\documentclass[apj]{emulateapj}

\usepackage{graphicx}
\usepackage[colorlinks=true,linkcolor=blue,linktoc=page,citecolor=blue,urlcolor=red]{hyperref}
\newcommand{\targetstar}{AS 209}

\shorttitle{Grain Growth in \targetstar\ Disk}
\shortauthors{P\'erez et al.}

\begin{document}

\title{Constraints on the Radial Variation of Grain Growth \\
in the \targetstar\ Circumstellar Disk}

\author{Laura M. P\'erez\altaffilmark{1},
John M. Carpenter\altaffilmark{1}, 
Claire J. Chandler\altaffilmark{2}, 
Andrea Isella\altaffilmark{1}, 
Sean M. Andrews\altaffilmark{3},
Luca Ricci\altaffilmark{1}, 
Nuria Calvet\altaffilmark{4},
Stuartt A. Corder\altaffilmark{5}, 
Adam T. Deller\altaffilmark{6}, 
Cornelis P. Dullemond\altaffilmark{7}, 
Jane S. Greaves\altaffilmark{8},
Robert J. Harris\altaffilmark{3},
Thomas Henning\altaffilmark{9},
Woojin Kwon\altaffilmark{10},
Joseph Lazio\altaffilmark{11},
Hendrik Linz\altaffilmark{7},
Lee G. Mundy\altaffilmark{12}, 
Anneila I. Sargent\altaffilmark{1},
Shaye Storm\altaffilmark{12}, 
Leonardo Testi\altaffilmark{13,14},
David J. Wilner\altaffilmark{3}
}
\altaffiltext{1}{California Institute of Technology, 1200 East California Blvd, Pasadena, CA 91125, USA}
\altaffiltext{2}{National Radio Astronomy Observatory, P.O. Box O, Socorro NM 87801, USA}
\altaffiltext{3}{Harvard-Smithsonian Center for Astrophysics, 60 Garden Street, Cambridge, MA 02138, USA}
\altaffiltext{4}{University of Michigan, 830 Dennison Building, 500 Church Street, Ann Arbor, MI 48109, USA}
\altaffiltext{5}{Joint ALMA Observatory, Av. Alonso de C\'ordova 3107, Vitacura, Santiago, Chile}
\altaffiltext{6}{The Netherlands Institute for Radio Astronomy (ASTRON), 7990-AA Dwingeloo, The Netherlands}
\altaffiltext{7}{Heidelberg University, Center for Astronomy, Albert Ueberle Str 2, Heidelberg, Germany}
\altaffiltext{8}{University of St. Andrews, Physics and Astronomy, North Haugh, St Andrews KY16 9SS}
\altaffiltext{9}{Max-Planck-Institut f\"{u}r Astronomie, K\"{o}nigstuhl 17, 69117 Heidelberg, Germany}
\altaffiltext{10}{University of Illinois, 1002 West Green St., Urbana, IL 61801, USA}
\altaffiltext{11}{Jet Propulsion Laboratory, California Institute of Technology, 4800 Oak Grove Dr, Pasadena, CA  91106}
\altaffiltext{12}{University of Maryland, College Park, MD 20742, USA}
\altaffiltext{13}{European Southern Observatory, Karl Schwarzschild str. 2, 85748 Garching, Germany}
\altaffiltext{14}{INAF-Osservatorio Astrofisico di Arcetri, Largo E. Fermi 5, 50125 Firenze, Italy}

\begin{abstract}


\noindent We present dust continuum observations of the protoplanetary disk surrounding the pre-main sequence star \targetstar, spanning more than an order of magnitude in wavelength from 0.88 to 9.8 mm. 
The disk was observed with sub-arcsecond angular resolution ($0.2''-0.5''$) to investigate radial variations in its dust properties.
At longer wavelengths, the disk emission structure is notably more compact, providing model-independent evidence for changes in the grain properties across the disk.
We find that physical models which reproduce the disk emission require a radial dependence of the dust opacity $\kappa_{\nu}$.
Assuming that the observed wavelength-dependent structure can be attributed to radial variations in the dust opacity spectral index ($\beta$), we find that $\beta(R)$ increases from $\beta<0.5$ at $\sim20$~AU to $\beta>1.5$ for $R\gtrsim80$~AU, inconsistent with a constant value of $\beta$ across the disk (at the 10$\sigma$ level).
Furthermore, if radial variations of $\kappa_{\nu}$ are caused by particle growth, we find that the maximum size of the particle-size distribution ($a_{max}$) increases from sub-millimeter-sized grains in the outer disk ($R\gtrsim70$~AU) to millimeter and centimeter-sized grains in the inner disk regions ($R\lesssim70$~AU). 
We compare our observational constraint on $a_{max}(R)$ with predictions from physical models of dust evolution in proto-planetary disks. For the dust composition and particle-size distribution investigated here, our observational constraints on $a_{max}(R)$ are consistent with models where the maximum grain size is limited by radial drift. 

\end{abstract}


\section{Introduction}

The growth of sub-micron-sized dust grains into millimeter and centimeter-sized particles is a fundamental component of the planet formation process \citep{2000Beckwith,2007Natta}. 
Grain growth directly affects the optical properties of dust particles \citep{1996Henning,2001DAlessio}, with composition and temperature having a lesser impact \citep{1997Henning,2003Semenov,2005Boudet}.
At long wavelengths, the slope $\alpha$ of the spectral energy distribution (SED, $\nu S_{\nu}\propto\nu^{\alpha+1}$) can be related to the spectral index of the dust opacity $\beta$ ($\kappa_{\nu}\propto\nu^{\beta}$), in particular $\beta=\alpha-2$ for optically thin warm dust \citep{1991Beckwith,1993Miyake}. 
The millimeter dust opacity slope for small interstellar medium (ISM) dust grains is $\beta_{\textrm{\footnotesize{ISM}}}\sim1.7$ \citep{2001Li_Draine}. A smaller value of $\beta$ is measured if larger millimeter-sized grains are present \citep{2001DAlessio,2006Draine}.

Observations of dust in disks from sub-mm to cm wavelengths \citep{2001Testi,2002Calvet,2003Testi,2004Natta,2005Wilner,2006Rodmann,2009Lommen,2010Ricci_a,2010Ricci_b} have provided strong evidence for grain growth in disks. These results, obtained mainly from spatially unresolved observations, assumed a constant dust opacity (i.e. constant $\beta$) throughout the disk. However, theoretical models of dust transport, fragmentation, and size evolution in proto-planetary disks predict that the average size of grains varies with the distance to the central star \citep[hereafter referred to as radius, ][]{2005DullemondDominik,2010Birnstiel}. Hence, radial variations in the dust opacity slope are expected.

The increased sensitivity and angular resolution of current interferometers has enabled the first studies of radial variations of grain growth within circumstellar disks. Observations of disks in Taurus at 1.3 and 3 mm with CARMA\footnote{Combined Array for Research in Millimeter-wave Astronomy} \citep{2010Isella} and PdBI\footnote{Plateau de Bure Interferometer} \citep{2011Guilloteau}, have constrained $\beta$ as a function of radius. However, these results are only sensitive to radial variations larger than $\Delta\beta\sim0.6-0.7$ at $3\sigma$ in the best cases, limited by the small wavelength separation between the two bands being observed. Increased wavelength coverage has been obtained by adding cm-wave observations \citep{2011Banzatti}, but the low signal-to-noise ratio, before the Karl G. Jansky Very Large Array (VLA) upgrade, made such studies challenging. To significantly improve these constraints, and effectively compare them to theoretical predictions of grain growth, increased wavelength coverage and enhanced sensitivity are required. 

This letter presents the first results of a program that combines spatially resolved observations from sub-millimeter to centimeter wavelengths, to constrain radial variations of grain growth within circumstellar disks.
The pre-main sequence star \targetstar, also known as V1121 Oph, is a classical K5 T-Tauri star with high accretion rate \citep[$1.3\times 10^{-7}\:\textrm{M}_{\sun}\:\textrm{yr}^{-1}$,][]{2000Johns-Krull}. \targetstar\ is isolated from the main $\rho$-Ophiuchus cloud ($\alpha=16^h49^m15.3^s,\delta=-14^d22^m08.7^s$) at a distance of $131\pm50\:\textrm{pc}$ \citep{2007vanLeeuwen}. 
Resolved 0.88 mm continuum observations reveal a smooth radial density structure down to $0.3''\sim40\:\textrm{AU}$ \citep{2009Andrews,2010Andrews}. We obtained additional spatially resolved observations at $\lambda>0.88\:\textrm{mm}$, to study grain growth inside the \targetstar\ disk.

\section{\targetstar\ Observations}

Submillimeter Array (SMA) observations at 0.88 mm were obtained between May 2006 and June 2007, in C, E, and V configurations ($\sim0.3''$ resolution), with total continuum bandwidth of 4~GHz. Its integrated flux density is $577\pm60\:\textrm{mJy}$ at  0.88 mm. \cite{2009Andrews} presented a detailed discussion of the observations and data reduction. 

CARMA observations at 2.8 mm were obtained between December 2009 and March 2010, in C, B, and A configurations spanning baselines of 30-1800~m ($\sim0.5''$ resolution), with total continuum bandwidth of 8~GHz. Its integrated flux density is $40\pm6\:\textrm{mJy}$ at 2.8 mm.  Atmospheric conditions were good, with zenith opacities $\tau_{\footnotesize\textrm{230 GHz}}\le0.15$, except for B configuration observations where $\tau_{\footnotesize\textrm{230 GHz}}\sim0.4$. Data were calibrated using the Multichannel Image Reconstruction, Image Analysis and Display (MIRIAD) software package \citep{1995Sault}.

VLA observations at 8.0 and 9.8 mm (Ka-band) were obtained between Jan-May 2011 (project AC982), in CnB and BnA configurations, spanning baselines of 350~m to 36.4~km ($\sim0.2''$ resolution). Two 1~GHz basebands centered at 30.5 and 37.5~GHz provided 2~GHz of continuum bandwidth. At 8.0 and 9.8 mm the integrated flux density of \targetstar\ is $1.2\pm0.1$ and $0.7\pm0.1$~mJy respectively. Data were calibrated using the Common Astronomy Software Applications (CASA). 

For all datasets, quasar observations interspersed between science target observations were used to calibrate the complex antenna gains. Absolute flux calibration was obtained by observing Neptune for CARMA, Uranus and Titan for SMA, and 3C286 for VLA. For each telescope, calibrated visibilities coming from different configurations were corrected for proper motion \citep{2007vanLeeuwen} and compared over the spatial frequencies where these configurations overlap; these were found to agree within the absolute flux scale uncertainty ($\sim10-15\%$). Fourier inversion of the visibilities and cleaning of the dirty image to produce dust emission map, were performed using CASA.

\begin{figure}[!t]
\begin{center}
\includegraphics[scale=0.42]{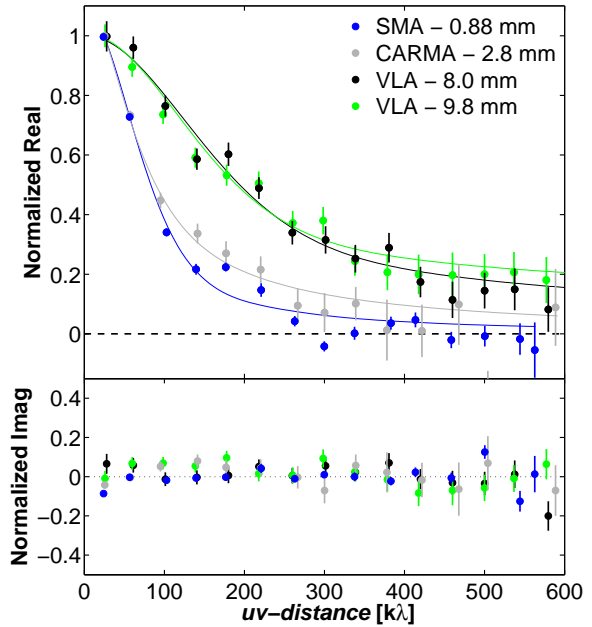}
\caption{Real and imaginary part of the bin-averaged visibilities from \targetstar\ as a function of spatial frequency (\emph{uv}-distance). 
Disk visibilities have been normalized by the measured flux density between \emph{uv}-distances of $0-40~[k\lambda]$, and deprojected assuming PA$=86^{\circ}$, $i=38^{\circ}$ \citep[][]{2009Andrews}. Filled circles and errorbars: correlated emission observed at different wavelengths, continuous lines: best-fit at each wavelength.}
\label{amp_vs_uvdist}
\end{center}
\end{figure}

\section{Observational Results}

Figure~\ref{amp_vs_uvdist} presents the normalized Real/Imaginary part of the correlated emission from \targetstar\, as a function of spatial frequency (\emph{uv}-distance), for each wavelength. 
The real part of the visibility will be constant with increasing spatial frequency for an unresolved point-source.
In contrast, for \targetstar\ we measure a declining profile with increasing {uv}-distance, indicative of spatially resolved emission at all wavelengths.
Furthermore, short-wavelength emission (0.88, 2.8 mm) diminishes faster than long-wavelength emission (8.0, 9.8 mm);
therefore, the structure at long-wavelengths is more compact than at short-wavelengths.

Before interpreting these observational results in terms of grain properties variations, we explore other explanations for the differing amplitude profiles: (1) seeing affecting high frequency observations would have to be $>1.2''$ at $0.88~$mm \citep[$\times10$ larger than observed seeing;][]{2009Andrews}, (2) optically thick emission at short wavelengths: modeling of these observations (\S\ref{ModelingResults}) indicates $\tau<1$ over the scales studied. Consequently, we conclude that the long-wavelength emission comes from a smaller region than the short-wavelength emission.

\begin{figure*}[!t]
\begin{center}
\includegraphics[scale=1.0]{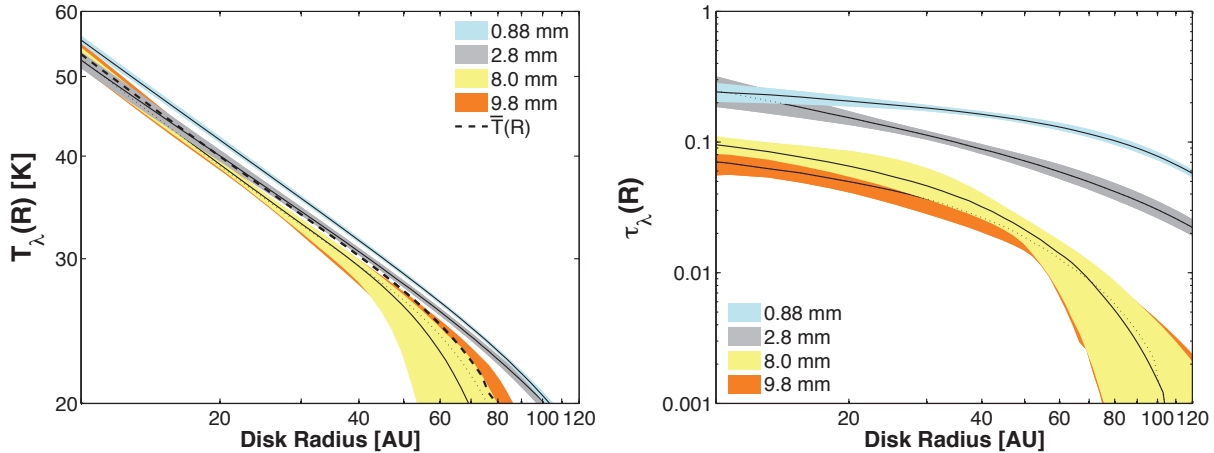}
\caption{\emph{Left:} $T_{\lambda}(R)$ inferred from separate modeling of our multi-wavelength observations, assuming a constant dust opacity with radius. \emph{Right:} Optical depth $\tau_{\lambda}(R)=\kappa_{\lambda}\times\Sigma_{\lambda}(R)$ inferred from separate modeling of multi-wavelength observations, assuming a radially constant $\kappa_{\lambda}$.
Colored regions: $3\sigma$ confidence interval constrained by our observations, continuous line: best-fit model, dashed line on left panel: average temperature profile. The different $\Sigma(R)$ and $T(R)$ profiles for each wavelength portray a varying dust opacity with radius, and because of this, none of them is the true surface density and temperature profile of the disk.}
\label{SigmaxT}
\end{center}
\end{figure*}

Before proceeding, we account for continuum emission not arising from dust, such as free-free emission from photo-evaporative stellar winds.
Such emission originates near the star \citep[within 10~AU for TW~Hya,][]{2011Pascucci}, appearing as a point-source at the resolution of our observations. 
Indeed, in addition to the resolved emission detected with VLA, we detect a point-like component (at {\it uv}-distances $>900\:\textrm{k}\lambda$) with a flux density of $0.08\pm0.02$~mJy (8 mm) and $0.10\pm0.02$~mJy (9.8 mm), corresponding to $\sim7\%$ and $\sim14\%$ of the total emission at these wavelengths.
However, VLA observations at 5.2~cm from the Disks@EVLA collaboration\footnote{https://safe.nrao.edu/evla/disks/} (Chandler et al. in prep) place a $3\sigma$ upper limit of $72\:\mu$Jy, suggesting that the point-like emission at 8 and 9.8 mm may be dust emission. Nonetheless, we conservatively include  this point-like emission as a free-free component in models described next.

\section{Modeling Results}
\label{ModelingResults}

Observations were analyzed using the disk emission model described in \cite{2009Isella}. The disk structure follows the two-layer disk approximation \citep{1997Chiang}, defined by the boundary between the disk surface layer (optically thin to stellar radiation) and the disk interior (optically thick to stellar emission). 
The disk is passively heated by \targetstar\ \citep[$L_{\star}=1.5\:L_{\sun},T_{\star}=4250\:$K, $M_{\star}=0.9\:M_{\sun}$,][]{2009Andrews}. Dust in the disk surface absorbs the stellar radiation and re-radiates this energy at longer wavelengths. Because this hot dust layer will be optically thin to its own thermal radiation, half of the surface layer emission will be radiated outwards while the rest will be radiated towards the disk, heating up its interior.

For the density structure, we employ the self-similar solution for a viscous accretion disk \citep{1974LBP}
\begin{eqnarray}
\Sigma(R) =&&\Sigma_t\left(\frac{R_t}{R}\right)^{\gamma} \nonumber \\ 
  &&\times\exp\left[ -\frac{1}{2(2-\gamma)}\left[\left(\frac{R}{R_t}\right)^{(2-\gamma)} -1\right]\right]
\end{eqnarray}
where $\Sigma_t$ corresponds to the surface density at radius $R_t$. This $\Sigma(R)$ prescription  behaves as a power-law for small radii, decreasing exponentially at large radii.

We compute the dust opacity $\kappa_{\nu}$ using Mie theory, assuming a population of compact spherical grains in a power-law size distribution, $n(a)\propto\:a^{-q}$, for $a_{min}<a<a_{max}$ with $a_{min}=0.01\:\mu$m. 
Optical constants for astrophysical grain components were obtained from \cite{2003Semenov}, \cite{1996Zubko}, and \cite{1984Warren}. 
Fractional abundances were adopted from \citet[][in volume: $8\%$ silicates, 30\% organics, 62\% water ice]{1994Pollack}. 
Recent revisions of the solar oxygen abundance \citep{2009Asplund} would reduce the amount of water-ice. However, the effect on the disk modeling is minimal, with a difference in the disk structure constraints well within their uncertainties as shown by \citet{2010Isella}.

The best-fit model to a single wavelength observation is found through $\chi^2$ minimization with 3 free parameters that describe the surface density ($\Sigma_t,R_t,\gamma$), since the disk geometry is fixed based on molecular line observations \citep[$i=38^{\circ}$, PA$=86^{\circ}$][]{2009Andrews}. The $\chi^2$ probability distribution is sampled by varying the free parameters, generating a model state and comparing this state to the data in Fourier space, following the Markov Chain Monte Carlo (MCMC) method described in \cite{2009Isella}.

For this initial model fit, we select the maximum grain size ($a_{max}$) and particle-size distribution slope ($q$) that best fit the unresolved SED from sub-mm to cm wavelengths. For \targetstar, these correspond to $q=3.5$, $a_{max}=1.3\;\textrm{mm}$, resulting in $\beta=1.0$ between 0.88 and 8.0 mm. Then, assuming a \emph{constant dust opacity} throughout the disk, each wavelength is fitted \emph{separately} with the aforementioned physical model. We obtain best-fit $\Sigma(R)$ and $T(R)$ profiles, that will be different for each wavelength if the assumption of a radially constant dust opacity is not valid \citep{2010Isella}.

\begin{figure*}[!t]
\begin{center}
%
%
\hspace{-0.3cm}\includegraphics[scale=1.24]{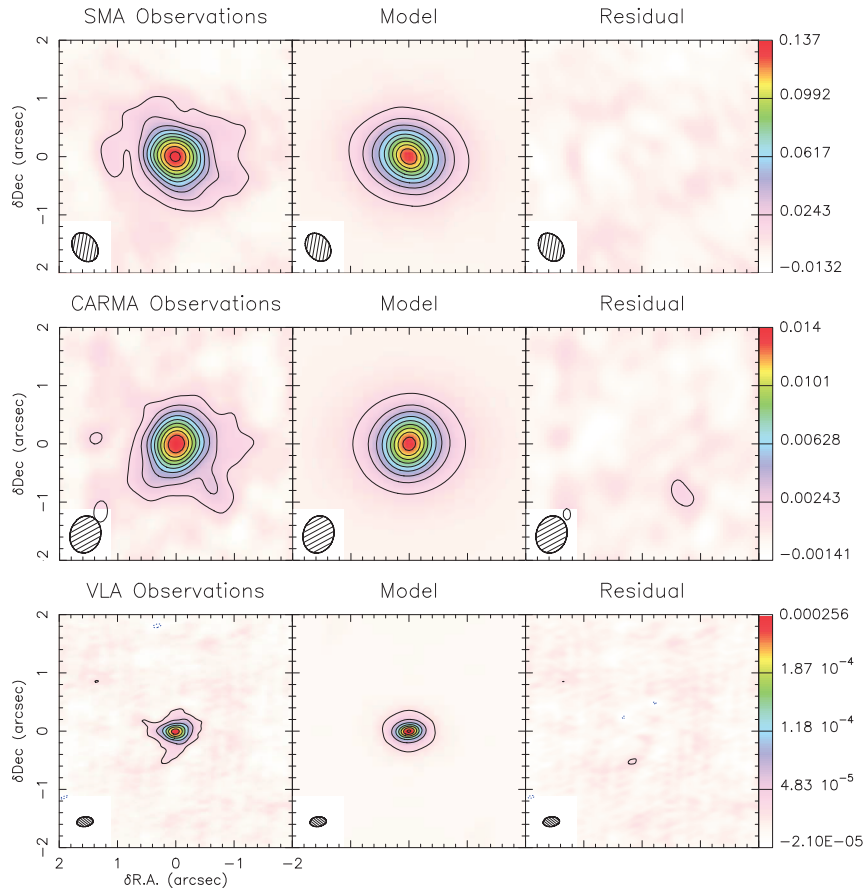}
\caption{Continuum emission towards \targetstar, observed at 0.88 mm (top panels), 2.8 mm (middle panels) and 8.0 and 9.8 mm (combined through multi-frequency synthesis, bottom panels). Each observation, accompanied by the best-fit  disk emission and a residual map obtained by subtracting the best-fit model from the observations, used Briggs weighting with robust $=0.7$ (SMA, CARMA), while VLA data used natural weighting. Contours start at $-3\sigma$, stepping by $3\sigma$ (CARMA, SMA) and $6\sigma$ (VLA), where  $\sigma$ is the RMS noise on each map: $\sigma_{\tiny\textrm{SMA}}=4.4$~mJy/beam, $\sigma_{\tiny\textrm{CARMA}}=0.47$~mJy/beam, $\sigma_{\tiny\textrm{VLA}}=0.01$~mJy/beam.}
\label{maps_model_res}
\end{center}
\end{figure*}

Figure~\ref{SigmaxT} shows the best-fit and $3\sigma$ constraints on $T(R)$ and the disk optical depth $\tau_{\lambda}(R)=\kappa_{\lambda}\Sigma(R)$, inferred from modeling each observation separately. 
As can be seen in this figure, a differing $\Sigma(R)$ and $T(R)$ was inferred for each wavelength when $\kappa_{\lambda}$ is assumed to be constant with radius. We constrain $R_t$ to be large for the short-wavelength emission ($R_t=61\pm2$~AU from 0.88 mm and $R_t=60\pm6$~AU from 2.8 mm), compared to $R_t$ inferred at long-wavelengths ($R_t=24\pm3$~AU from 8 mm, $R_t=26\pm4$~AU from 9.8 mm). We note that the derived temperature profiles differ by $<4$~K for $R<70$~AU, and by $\lesssim10$~K for $R>70$~AU.
The best-fit model visibility profiles at each wavelength are shown as continuous lines in Figure~\ref{amp_vs_uvdist}. Maps of the observed emission, with corresponding best-fit model and residual emission maps are shown in Figure~\ref{maps_model_res}.

Since the disk is optically thin (right panel, Figure~\ref{SigmaxT}), the observed emission will depend directly upon the dust opacity and disk mass: $S_{\lambda}\propto\kappa_{\lambda}\times\Sigma\times\:B_{\lambda}(T)$. Hence, our observations constrain the product $\kappa_{\lambda}\times\Sigma\times\:B_{\lambda}(T)$, where all these physical parameters ($\kappa_{\lambda},\Sigma,T$) depend on the radius, but only the dust opacity depends on the observed wavelength. 
The modeling presented above, assuming a constant dust opacity with radius and fitting each wavelength separately, has resulted in a wavelength-dependent $\Sigma_{\lambda}(R)$ and $T_{\lambda}(R)$, which is unphysical but telling: the assumption of a constant $\kappa_{\lambda}$ with radius is likely incorrect. Radial variations of the dust opacity are required to reconcile the differing visibility profiles (Real part vs. $uv$-distance, Figure~\ref{amp_vs_uvdist}) and the wavelength-dependent $\Sigma_{\lambda}(R)$ and $T_{\lambda}(R)$ (Figure~\ref{SigmaxT}).

\begin{figure*}[!t]
\begin{center}
\includegraphics[scale=1.0]{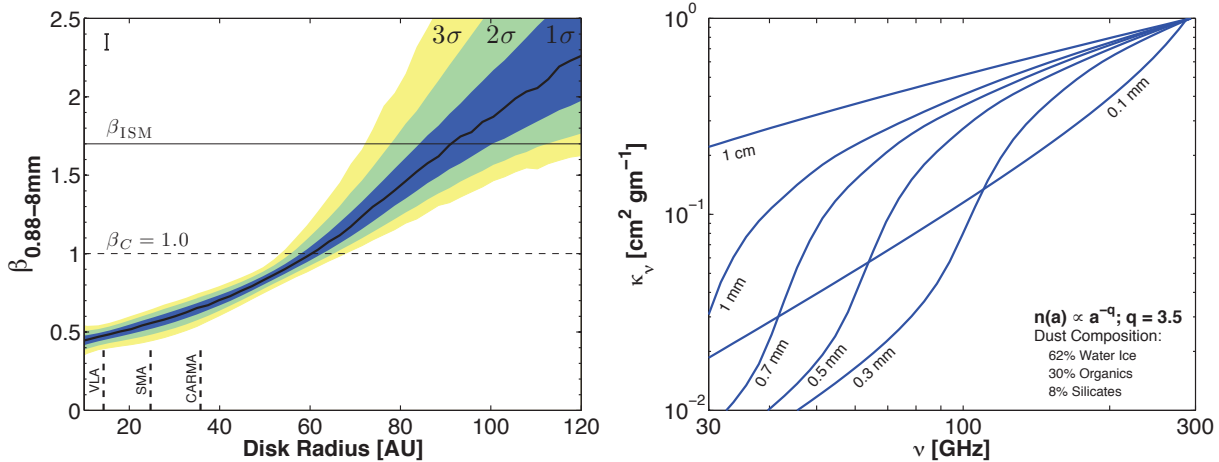}
\caption{{\it Left:} Dust opacity spectral slope, $\beta$, vs. radius, inferred from multi-wavelength observations of the \targetstar\ disk. Black line: best-fit $\beta(R)$, colored areas: confidence interval constrained by our observations. Vertical dashed-lines indicate the spatial resolution of our observations, errorbar in top-left corner indicates additional systematic uncertainty on $\beta(R)$ arising from amplitude calibration uncertainty. {\it Right:} Dust opacity (normalized at 300~GHz) for $a_{max}$ between 0.1-10~cm. Note that the power-law assumption, $\kappa_{\nu}\propto\nu^{\beta}$, breaks down for (sub-)mm-sized grains.}
\label{Fig_beta}
\end{center}
\end{figure*}

\section{Discussion}

\subsection{Radial variations of $\beta$}

Expanding on the work presented in \citet{2010Isella}, if the \emph{true} but unknown disk physical quantities are $\overline\kappa_{\lambda}(R)$, $\overline\Sigma(R)$, and $\overline T(R)$, then at each wavelength
\begin{equation}
\kappa_{\lambda}\:\Sigma_{\lambda}(R)\:B_{\lambda}(T_{\lambda}(R))=\overline\kappa_{\lambda}(R)\:\overline\Sigma(R)\:B_{\lambda}(\overline T(R))
\label{equality}
\end{equation}
where the right-hand side of this equation corresponds to the true disk properties, and the left-hand side encompasses the constraints found from our multi-wavelength observations. 
Taking the ratio of equation~\ref{equality} for $\lambda_1\ne\lambda_2$,  and assuming that at long wavelengths $\kappa_{\lambda}\propto\lambda^{-\beta}$, we obtain
\begin{eqnarray}
\Delta\beta(R)=\log^{-1}\left[\frac{\lambda_1}{\lambda_2}\right]\times\log\left[\frac{\Sigma_{\lambda_1}B_{\lambda_1}(T_{\lambda_1})/B_{\lambda_1}(\overline T)}{\Sigma_{\lambda_2}B_{\lambda_2}(T_{\lambda_2})/B_{\lambda_2}(\overline T)}\right] 
\label{Deltabeta}
\end{eqnarray}
where $\Sigma_{\lambda}$, $T_{\lambda}$, and $\overline T$ depend on the radius $R$.
Equation~\ref{Deltabeta} is a useful prescription to constrain radial variations of the opacity spectral slope ($\Delta\beta(R)=\beta(R)-\beta_C$) in dual-wavelength observations. Here $\beta_C$ corresponds to the spectral slope of the assumed radially constant dust opacity.

In the Rayleigh-Jeans regime (when $hc/\lambda\ll~k_BT$, $k_B$: Boltzmann constant), knowledge of $\overline T(R)$ to derive $\Delta\beta(R)$ is unnecessary since $B_{\lambda}(T)\propto T$. However, in the cold outer disk and at short wavelengths, the Rayleigh-Jeans assumption is ill-founded (e.g. $h\nu\sim~k_BT$ for $T=20$~K at 0.88 mm). Consequently, we require an estimate of $\overline T(R)$. Since the temperature profiles inferred from each wavelength ($T_{\lambda}(R)$) are not very different, we average them  to deduce $\overline T(R)$ (dashed-line on Figure~\ref{SigmaxT}).

	Equation~\ref{Deltabeta} shows that in logarithmic space $\Delta\beta(R)$ is the slope of a line that goes through points \{$x=\log(\lambda)$, $y=\log[\Sigma_{\lambda}B_{\lambda}(T_{\lambda})/B_{\lambda}(\overline T)]$\}, with $y$ evaluated at radius~$R$. 
	Hence, to measure $\Delta\beta(R)$ we find the slope of a straight line -- in a least-square sense -- that fits our constraints on the product $\Sigma_{\lambda}\times\:B_{\lambda}(T_{\lambda})/B_{\lambda}(\overline T)$ from our set of multi-wavelength observations. 
	To set confidence intervals on $\Delta\beta(R)$, we make use of the Bayesian approach upon which our modeling is based. 
	Since the result of the MCMC algorithm is a fully sampled posterior probability distribution function (PDF) for all the model parameters,  we construct a PDF of the product $\Sigma_{\lambda}\times\:B_{\lambda}(T_{\lambda})/B_{\lambda}(\overline T)$ at each radius $R$ and for each  wavelength. 	
	Random samples of these PDFs are taken at each wavelength, the slope of the line through points \{$x=\log(\lambda)$, $y=\log[\Sigma_{\lambda}\times B_{\lambda}(T_{\lambda})/B_{\lambda}(\overline T)]$\} at a radius $R$, is computed. This slope corresponds to one random sampling of $\Delta\beta(R)$. Hence, the PDF for $\Delta\beta(R)$ is constructed by performing a large number of random samples. The peak of this PDF is the best-fit value of $\Delta\beta$ at radius $R$.
	Confidence intervals are derived from the region of the PDF that contains 68.3\%,~95.5\%~and~99.7\% of all samples at equal probability ($1\sigma,\:2\sigma$,~and~$3\sigma$).

	Figure~\ref{Fig_beta} presents the constraints on the radial variation of $\beta$ obtained from our multi-wavelength observations. 
	The values of $\beta$ allowed by our observations are significantly different than $\beta_{\textrm{ISM}}\sim1.7$, for $R\lesssim70$~AU. 
	Furthermore, we find a gradient on $\beta(R)$ inconsistent with a constant value at the 10$\sigma$ level.

\subsection{Radial variations of $a_{max}$}

To derive equation~\ref{Deltabeta}, the assumption of $\kappa_{\lambda}\propto\lambda^{-\beta}$ must be satisfied. We caution that  for $a_{max}\sim0.1$-1 mm this approximation may break down \citep{2006Draine}, as illustrated in Figure~\ref{Fig_beta} (right). Therefore, rather than inferring $a_{max}(R)$ from $\beta(R)$, we constrain it directly by fitting a specific dust opacity $\overline\kappa_{\lambda}$ to the constraints on the product $\kappa_{\lambda}\times\Sigma_{\lambda}\times B_{\lambda}(T_{\lambda})$ at each radius.
With a knowledge of $\overline T(R)$, and for a fixed set of dust properties (composition and grain-size distribution), we estimate the values of $a_{max}$ and $\overline\Sigma$ that satisfy equation \ref{equality}, now written as
\begin{equation}
\kappa_{\lambda}\:\Sigma_{\lambda}(R)\:\frac{B_{\lambda}(T_{\lambda}(R))}{B_{\lambda}(\overline T(R))} =\overline\kappa_{\lambda}(a_{max}(R))\:\overline\Sigma(R)
\label{new_equality}
\end{equation}
where the right-hand side corresponds to our model (with parameters $a_{max}$ and $\overline\Sigma$), and the left-hand side has been constrained by our multi-wavelength observations (i.e. we have a PDF for the product $\kappa_{\lambda}\times\Sigma_{\lambda}(R)\times\frac{B_{\lambda}(T_{\lambda}(R))}{B_{\lambda}(\overline T(R))}$).

At each radius R, we constructed a two dimensional grid of parameters $\{a_{max},\overline\Sigma\}$. At each point in the grid, we compute the product $\overline\kappa_{\lambda}(a_{max})\times\overline\Sigma$ at each wavelength. We then find the probability that such measurement will have occurred (given our observational constraints on the left-hand side of equation~\ref{new_equality}), and construct the likelihood function of the parameters $\{a_{max},\overline\Sigma\}$. Best-fit values for $a_{max}$ and $\overline\Sigma$ are found by maximizing the likelihood, confidence intervals are obtained from the marginalized likelihoods.

\begin{figure*}[!t]
\begin{center}
\leavevmode
\includegraphics[scale=1.0]{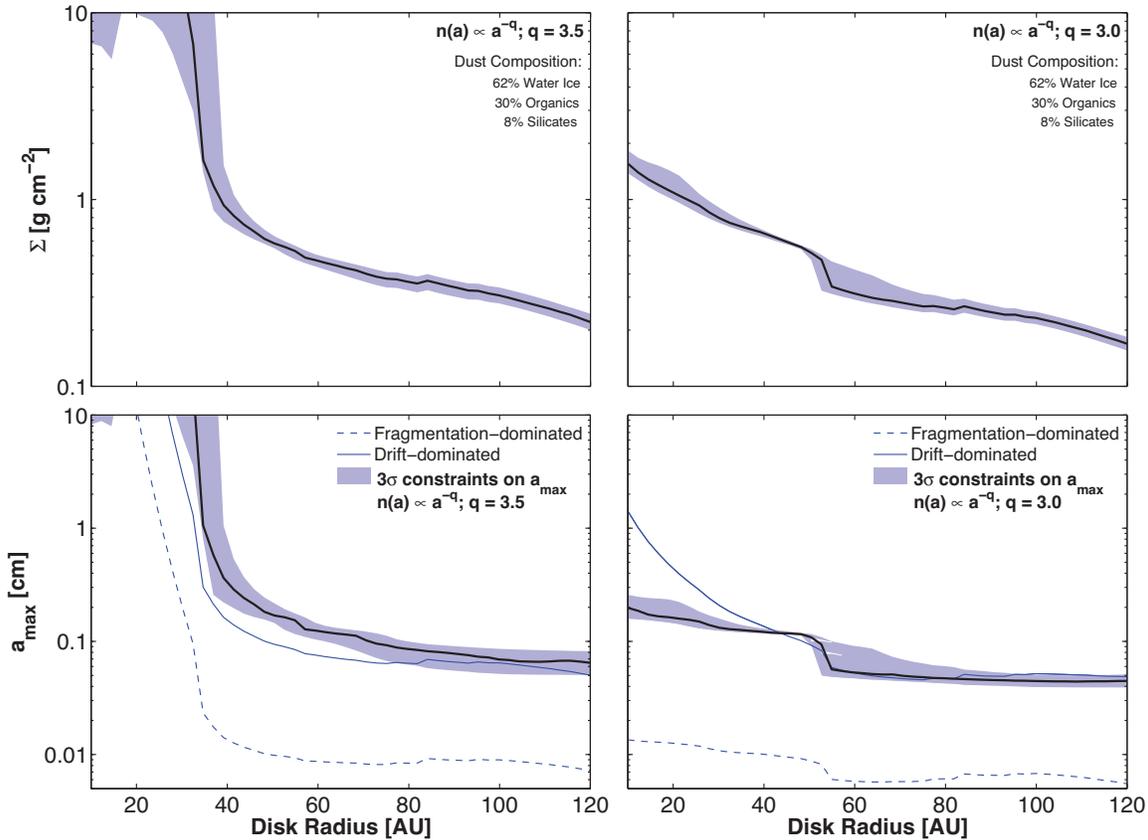}
\caption{Surface density (top) and maximum grain size as a function of radius (bottom), for two grain size distributions: $q=3.5$ (left), $q=3.0$ (right). Black line: best-fit, shaded region: $3\sigma$ confidence interval. Assumed grain properties are specified in figure legend. We compare our observational constraints with theoretical grain evolution models \citep{2012Birnstiel}, that include fragmentation and radial drift.}
\label{Fig_amax}
\end{center}
\end{figure*}

Figure~\ref{Fig_amax} presents our constraints on $a_{max}(R)$ and $\overline\Sigma(R)$ for two representative values of $q$. 
The same composition presented in \S\ref{ModelingResults} is assumed, however a different dust mixture will influence the derived $a_{max}(R)$. For a composition that includes updated oxygen abundances \citep{2009Asplund}, a smaller $a_{max}(R)$ is inferred ($\times2$ smaller), well within the uncertainties of this derivation given other unknown parameters (e.g. particle-size distribution slope). 
Across the disk, grains have grown at least up to $\sim0.5$ mm, with small grains present in the outer disk and large grains in the inner disk.

We compare our observational constraints with theoretical models of grain growth, employing the approximations presented in \cite{2012Birnstiel} for the evolution of $a_{max}$ with radius. Since the true $\overline\Sigma(R)$ profile constrained by our observations depends on the assumed value of $q$ (see Figure~\ref{Fig_amax}), these theoretical prescriptions will depend on $q$ as well.
Our observational constraints on $a_{max}$ are consistent with a \emph{radial drift limited} grain population, where the head-wind felt by dust particles makes them spiral towards the star. A \emph{fragmentation dominated} population, where the turbulent relative motion of particles causes collisions that either grow or fragment these grains, seems incompatible for standard values of the turbulence parameter \citep[$\alpha_t=0.01$,][]{1973Shakura}, fragmentation threshold velocity \citep[$u_t=10$ m/s,][]{2008Blum_Wurm}, and 100:1 gas-to-dust ratio. These curves represent barriers that prevent further size increase, hence, the smaller of the two is considered the upper-limit to growth. However, the parameters that go into deriving the fragmentation limit are very uncertain, while the physics and parameters in radial drift are better established. Either no fragmentation barrier exists (for a low-turbulence disk, $\alpha_t\lesssim0.001$, the maximum collision speed never reaches the fragmentation threshold velocity in \targetstar, making fragmentation impossible) or the gas-to-dust ratio is $\sim\times10$ larger, allowing for the fragmentation-limited barrier to be consistent with our $a_{max}(R)$ constraint.
It remains to be demonstrated whether these models reproduce the observed $\beta(R)$ profile and the millimeter fluxes. Such analysis requires numerical simulations starting from the disk parameters derived here and will be discussed in a future paper.

\subsection{Summary}

We have obtained multi-wavelength spatially-resolved observations at 0.88, 2.8, 8.0, and 9.8 mm of the \targetstar\ circumstellar disk. 
These observations reveal a wavelength-dependent structure, explained as radial variations of the dust opacity across the disk. 
We find a change in $\Delta\beta>1$ between the inner ($\sim20$~AU) and outer ($\sim120$~AU) disk, inconsistent with a constant $\beta$ value. This gradient in $\beta(R)$ implies that a significant change in the dust properties as a function of radius must exist. 
We interpret this gradient as a decrease in the maximum grain size with radius: going from several centimeters or more in the inner disk to sub-mm-sized grains in the outer disk.
When compared with theoretical models of dust size evolution \citep{2012Birnstiel}, we find that our observational constraint on $a_{max}(R)$ agrees with a radial-drift-dominated population, for reasonable values of the composition, particle-size distribution, and disk properties.

\acknowledgments
We acknowledge T. Birnstiel for useful discussions. A.I., J.M.C., L.M.P acknowledge support from NSF award AST-1109334.
The National Radio Astronomy Observatory is a facility of the National Science Foundation operated under cooperative agreement by Associated Universities, Inc.
Ongoing CARMA development and operations are supported by the National Science Foundation under a cooperative agreement, and by the CARMA partner universities.
The SMA is a joint project between the Smithsonian Astrophysical Observatory and the Academia Sinica Institute of Astronomy and Astrophysics, funded by the Smithsonian Institution and Academia Sinica.
Part of this research was carried out at the Jet Propulsion Laboratory, Caltech, under a contract with the National Aeronautics and Space Administration.

{\it Facilities:}\facility{CARMA}, \facility{SMA}, \facility{VLA}.



\begin{thebibliography}{}

\bibitem[Andrews et al.(2009)]{2009Andrews} Andrews, S.~M., Wilner, D.~J., Hughes, A.~M., Qi, C., \& Dullemond, C.~P.\ 2009, \apj, 700, 1502 

\bibitem[Andrews et al.(2010)]{2010Andrews} Andrews, S.~M., Wilner, D.~J., Hughes, A.~M., Qi, C., \& Dullemond, C.~P.\ 2010, \apj, 723, 1241 

\bibitem[Asplund et al.(2009)]{2009Asplund} Asplund, M., Grevesse, N., Sauval, A.~J., \& Scott, P.\ 2009, \araa, 47, 481 

\bibitem[Banzatti et al.(2011)]{2011Banzatti} Banzatti, A., Testi, L., Isella, A., et al.\ 2011, \aap, 525, A12 


\bibitem[Beckwith \& Sargent(1991)]{1991Beckwith} Beckwith, S.~V.~W., \& Sargent, A.~I.\ 1991, \apj, 381, 250 

\bibitem[Beckwith et al.(2000)]{2000Beckwith} Beckwith, S.~V.~W., Henning, T., \& Nakagawa, Y.\ 2000, Protostars and Planets IV, 533 

\bibitem[Birnstiel et al.(2010)]{2010Birnstiel} Birnstiel, T., Dullemond, C.~P., \& Brauer, F.\ 2010, \aap, 513, A79 

\bibitem[Birnstiel et al.(2012)]{2012Birnstiel} Birnstiel, T., Klahr, H., \& Ercolano, B.\ 2012, \aap, 539, A148 

\bibitem[Blum \& Wurm(2008)]{2008Blum_Wurm} Blum, J., \& Wurm, G.\ 2008, \araa, 46, 21 

\bibitem[Boudet et al.(2005)]{2005Boudet} Boudet, N., Mutschke, H., Nayral, C., et al.\ 2005, \apj, 633, 272 

\bibitem[Calvet et al.(2002)]{2002Calvet} Calvet, N., D'Alessio, P., Hartmann, L., et al.\ 2002, \apj, 568, 1008 

\bibitem[Chiang \& Goldreich(1997)]{1997Chiang} Chiang, E.~I., \& Goldreich, P.\ 1997, \apj, 490, 368 

\bibitem[D'Alessio et al.(2001)]{2001DAlessio} D'Alessio, P., Calvet, N., \& Hartmann, L.\ 2001, \apj, 553, 321 

\bibitem[Draine(2006)]{2006Draine} Draine, B.~T.\ 2006, \apj, 636, 1114 


\bibitem[Dullemond \& Dominik(2005)]{2005DullemondDominik} Dullemond, C.~P., \& Dominik, C.\ 2005, \aap, 434, 971 

\bibitem[Guilloteau et al.(2011)]{2011Guilloteau} Guilloteau, S., Dutrey, A., Pi{\'e}tu, V., \& Boehler, Y.\ 2011, \aap, 529, A105 

\bibitem[Henning \& Mutschke(1997)]{1997Henning} Henning, T., \& Mutschke, H.\ 1997, \aap, 327, 743 

\bibitem[Henning \& Stognienko(1996)]{1996Henning} Henning, T., \& Stognienko, R.\ 1996, \aap, 311, 291 

\bibitem[Isella et al.(2009)]{2009Isella} Isella, A., Carpenter, J.~M., \& Sargent, A.~I.\ 2009, \apj, 701, 260 

\bibitem[Isella et al.(2010)]{2010Isella} Isella, A., Carpenter, J.~M., \& Sargent, A.~I.\ 2010, \apj, 714, 1746 

\bibitem[Johns-Krull et al.(2000)]{2000Johns-Krull} Johns-Krull, C.~M., Valenti, J.~A., \& Linsky, J.~L.\ 2000, \apj, 539, 815 


\bibitem[Li \& Draine(2001)]{2001Li_Draine} Li, A., \& Draine, B.~T.\ 2001, \apj, 554, 778 

\bibitem[Lommen et al.(2009)]{2009Lommen} Lommen, D., Maddison, S.~T., Wright, C.~M., et al.\ 2009, \aap, 495, 869 

\bibitem[Lynden-Bell \& Pringle(1974)]{1974LBP} Lynden-Bell, D., \& Pringle, J.~E.\ 1974, \mnras, 168, 603 

\bibitem[Miyake \& Nakagawa(1993)]{1993Miyake} Miyake, K., \& Nakagawa, Y.\ 1993, Icarus, 106, 20 

\bibitem[Natta \& Testi(2004)]{2004Natta} Natta, A., \& Testi, L.\ 2004, Star Formation in the Interstellar Medium: In Honor of David Hollenbach, 323, 279 

\bibitem[Natta et al.(2007)]{2007Natta} Natta, A., Testi, L., Calvet, N., et al.\ 2007, Protostars and Planets V, 767 

\bibitem[Pascucci et al.(2011)]{2011Pascucci} Pascucci, I., Sterzik, M., Alexander, R.~D., et al.\ 2011, \apj, 736, 13 

\bibitem[Pollack et al.(1994)]{1994Pollack} Pollack, J.~B., Hollenbach, D., Beckwith, S., et al.\ 1994, \apj, 421, 615 

\bibitem[Ricci et al.(2010a)]{2010Ricci_a} Ricci, L., Testi, L., Natta, A., \& Brooks, K.~J.\ 2010, \aap, 521, A66 

\bibitem[Ricci et al.(2010b)]{2010Ricci_b} Ricci, L., Testi, L., Natta, A., et al.\ 2010, \aap, 512, A15 


\bibitem[Rodmann et al.(2006)]{2006Rodmann} Rodmann, J., Henning, T., Chandler, C.~J., Mundy, L.~G., \& Wilner, D.~J.\ 2006, \aap, 446, 211 
\bibitem[Sault et al.(1995)]{1995Sault} Sault, R.~J., Teuben, P.~J., \& Wright, M.~C.~H.\ 1995, Astronomical Data Analysis Software and Systems IV, 77, 433 

\bibitem[Semenov et al.(2003)]{2003Semenov} Semenov, D., Henning, T., Helling, C., Ilgner, M., \& Sedlmayr, E.\ 2003, \aap, 410, 611 

\bibitem[Shakura \& Sunyaev(1973)]{1973Shakura} Shakura, N.~I., \& Sunyaev, R.~A.\ 1973, \aap, 24, 337 

\bibitem[Testi et al.(2001)]{2001Testi} Testi, L., Natta, A., Shepherd, D.~S., \& Wilner, D.~J.\ 2001, \apj, 554, 1087 

\bibitem[Testi et al.(2003)]{2003Testi} Testi, L., Natta, A., Shepherd, D.~S., \& Wilner, D.~J.\ 2003, \aap, 403, 323 


\bibitem[van Leeuwen(2007)]{2007vanLeeuwen} van Leeuwen, F.\ 2007, \aap, 474, 653 

\bibitem[Warren(1984)]{1984Warren} Warren, S.~G.\ 1984, \ao, 23, 1206 


\bibitem[Wilner et al.(2005)]{2005Wilner} Wilner, D.~J., D'Alessio, P., Calvet, N., Claussen, M.~J., \& Hartmann, L.\ 2005, \apjl, 626, L109

\bibitem[Zubko et al.(1996)]{1996Zubko} Zubko, V.~G., Mennella, V., Colangeli, L., \& Bussoletti, E.\ 1996, \mnras, 282, 1321 

\end{thebibliography}
\end{document}